\documentclass[pra,twocolumn,superscriptaddress,showpacs,floatfix]{revtex4}
\usepackage{graphicx}
\usepackage{amssymb}
\usepackage{amsmath}
\usepackage{amstext}
\usepackage{latexsym}
\usepackage[matrix,frame,arrow]{xypic}

\newcommand{\beq}{\begin{equation}}
\newcommand{\eeq}{\end{equation}}
\newcommand{\barr}{\begin{eqnarray}}
\newcommand{\earr}{\end{eqnarray}}    
\newcommand{\ket}[1]{\left\vert#1\right\rangle}
\newcommand{\bra}[1]{\left\langle#1\right\vert}

\newcommand{\abs}[1]{\left\vert #1 \right\vert}

\newcommand{\fid}{\mathcal F}
\newcommand{\eps}{\varepsilon}

\begin{document}

\title{From perfect to fractal transmission in spin chains}
\author{Gabriele De Chiara}
\affiliation{ NEST- INFM \& Scuola Normale Superiore, Piazza dei
Cavalieri 7, I-56126 Pisa, Italy}
\homepage{www.qti.sns.it}
\author{Davide Rossini}
\affiliation{ NEST- INFM \& Scuola Normale Superiore, Piazza dei
Cavalieri 7, I-56126 Pisa, Italy}
\homepage{www.qti.sns.it}
\author{Simone Montangero}
\affiliation{ NEST- INFM \& Scuola Normale Superiore, Piazza dei
Cavalieri 7, I-56126 Pisa, Italy}
\homepage{www.qti.sns.it}
\author{Rosario Fazio}
\affiliation{ NEST- INFM \& Scuola Normale Superiore, Piazza dei
Cavalieri 7, I-56126 Pisa, Italy}
\homepage{www.qti.sns.it}

\date{\today}

\begin{abstract}
Perfect state transfer is possible in modulated spin chains
[Phys. Rev. Lett. {\bf 92}, 187902 (2004)],
imperfections however are likely to corrupt the state transfer. We study the 
robustness of this quantum communication protocol in the presence of disorder 
both in the exchange couplings between the spins and in the local magnetic 
field. The degradation of the fidelity can be suitably expressed, as a function
of the level of imperfection and the length of the chain, in a scaling form.
In addition the time signal of fidelity becomes fractal. We further characterize
the state transfer by analyzing the spectral properties of the
Hamiltonian of the spin chain.
\end{abstract}


\maketitle

\section{Introduction}
The ability to transfer a quantum state between distant parties 
is one of the basic requirements in many quantum information protocols, 
we mention for example quantum key distribution~\cite{gisin} or 
teleportation~\cite{teleportation}. A very successful area where the 
implementation of quantum state transmission has been realized is  
quantum optics. The carriers of information (photons) can be addressed 
and transmitted with high control and with a low level of decoherence.
Very recently, in view of the great potentialities of solid-state 
quantum information, attention is also focusing on the problem of the 
transfer of quantum information in a solid-state environment. A possible 
way to follow would be to properly design couplings between optical 
and solid-state systems~\cite{tian}. Alternatively one could also think 
to realize quantum channels using condensed-matter systems. 
In Ref.~\cite{bose} Bose has shown that a Heisenberg spin chain is able 
to act as a quantum channel over reasonable distance ($\sim 10^2$ lattice 
sites). Information capacities for this Heisenberg channel have been analyzed in 
Ref.~\cite{giovannetti}. In Ref.~\cite{Li} a slightly different scheme
has been proposed, in which the simple spin chain has been replaced
with an isotropic antiferromagnetic spin ladder.
A great advantage of these approaches is that state 
transfer occurs due to the interaction between the spins of 
the chain and no dynamical control is required (except for the preparation and the
detection of the state). Proposals to implement this scheme with superconducting
nanocircuits~\cite{romito,paternostro} have been already put forward and very 
likely these implementations can also be extended to other solid-state systems.

Perfect transfer can be achieved over arbitrary distances in spin
chains under many different hypotheses: by a proper choice of the modulation 
of the coupling strengths as suggested~\cite{christandl03}, 
if local measurements 
on the individual spins can be implemented~\cite{veer}, when communicating
parties have access to limited numbers of qubits in a spin ring~\cite{osborne03}
or by using several spin chains in parallel~\cite{burgarth}.
As in all situations in quantum information, the efficiency of such
protocols relies on the capability of isolating the experimental
setup from the external world (decoherence), and on the possibility to 
reduce all possible static imperfections~\cite{core,benenti}.
In the case of the protocols presented in Refs.~\cite{bose,christandl03}, where 
no control is needed on the system during the state transfer, that is, no 
dynamical control is applied, the coupling to the environment is supposed 
to be weak. It thus remains to be seen how the quantum channel is robust against
static imperfections which would be unavoidable (especially in solid-state 
implementations with engineered nanodevices). Particularly important to 
consider is the case when, in the ideal case, perfect state transfer is 
obtained.

In this paper we address the problem of the effects of 
static imperfections on the protocol presented in Ref.~\cite{christandl03},
which has already been experimentally implemented in Ref.~\cite{Zhang}
using a three qubit nuclear magnetic-resonance quantum computer. 
We study the sensitivity of the state transfer to random variations both
of the coupling between the spins and of an externally applied magnetic field. 
In view of the possible applications with solid-state systems, also 
the case of correlated disorder will be considered.
Similar questions for quantum computation protocols have been already 
analyzed in Refs.~\cite{core,benenti}. In that case the loss of efficiency of the protocol
was related to the appearance of quantum chaos in a quantum computer register. 
This relation has been characterized studying the level spacing 
statistics. Following  these lines we study the transition of
the level spacing statistics of the spin chain in the presence of static 
imperfections. Even though it is not possible to 
frame this problem with the random matrix theory 
\cite{rmt}, we show that the level spacing statistic is still a 
convenient tool to describe the system efficiency in performing 
the state transfer. 

The presence of static imperfections leads to another clear signature
of the modified properties of the spectrum in the fidelity. The degradation 
of the state transfer corresponds to the emergence of a fractal signal, 
i.e., the fidelity changes from a periodic function of time to a fractal 
time series. This behavior has the same origin as the one found in the probability 
densities of the quantum evolution in tight-binding 
lattices~\cite{berry,amanatidis}.

The paper is organized as follows. In Sec.~\ref{sec:model} we introduce
the model used throughout this work, we set up the notations, and we briefly 
review the quantum state transmission protocol of Refs.~\cite{bose,christandl03}. 
We then analyze the fidelity of the transferred state (Sec.~\ref{sec:stab}), and
the level spacing statistics of the Hamiltonian of the system (Sec.~\ref{sec:lss}). 
Interestingly, the dependence of the fidelity, as a function of the length of 
the chain and the level of disorder, obeys simple scaling laws.
In Sec.~\ref{sec:frac} we take a closer look at the behavior of the fidelity 
as a function of time. The presence of static imperfections leads 
to a fractal behavior of the time signal of the fidelity. In the 
same section, we relate the fractal dimension to the amount of disorder present 
in the chain. The last section is devoted to the conclusions. 

\section{Model} 
\label{sec:model}
The protocol introduced in Ref.\cite{bose}
enables quantum state transfer between two parties by means of a
spin chain: The state of the left-most qubit is transferred to
the right-most qubit after a given time (dictated by the dynamics of 
the chain). In Ref.\cite{christandl03} the approach is the same as in
Ref.\cite{bose}, the idea is to use a modulated chain whose Hamiltonian is given by  

\begin{equation}
	H=\sum_{k=1}^N B_k \, \sigma_k^z + \sum_{k=1}^{N-1}
	J_k \, (\sigma^x_k \sigma^x_{k+1} + \sigma^y_k \sigma^y_{k+1}).
\label{spinchain} 
\end{equation}
In Eq. (\ref{spinchain}) $N$ is the number of spins in the chain,
$\sigma_k^x$, $\sigma_k^y$, $\sigma_k^z$ are the Pauli operators
of the $k$th spin.
The parameters $B_k$ and $J_k$ are, respectively, the local magnetic field
and the exchange coupling constant. Both couplings depend on the position 
of the site (or of the link) on the chain. In order to achieve perfect 
state transfer, the system parameters are chosen to
be 
$$
	B_k=0, \qquad J_k = J \sqrt{k(N-k)} \, .
$$
The spin chain is initially (at time $t=0$) prepared in the state
\begin{equation} 
	\vert \Psi_0 (\vartheta,\varphi)\rangle = 
	\left(\cos\, \vartheta \ket 0 + \sin\, \vartheta \, e^{\imath \varphi} 
	\ket 1\right) \otimes \ket{0}^{\otimes (N-1)},
\label{icond}
\end{equation}
that is, the left-most spin is prepared in a given superposition of its two
levels while the others are in their ground state. 
\begin{figure}[htbp]
  	\begin{center}
     	\includegraphics[scale=0.35]{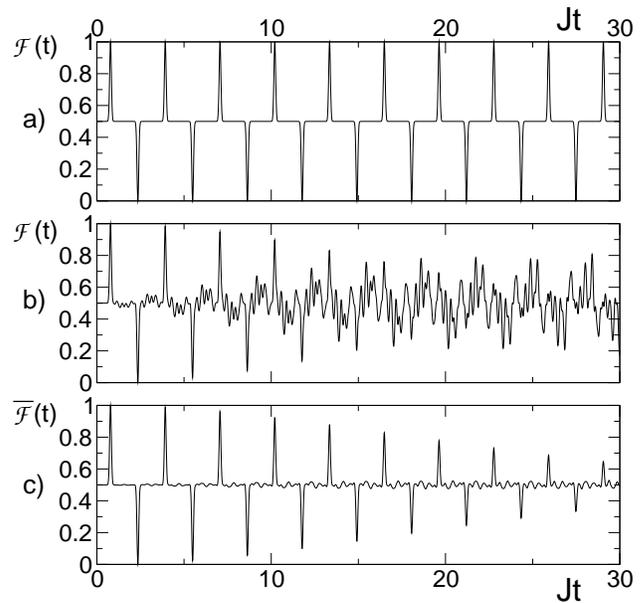}
  	\caption{Fidelity of the $N$th spin with $N=100$ as a function of
    	time. {\bf a)}~Without imperfections, $\eps_J=\eps_B=0$. {\bf
    	b)}~With imperfections $\eps_J=10^{-2} , \, \eps_B=0, N_{av}=1$.
	{\bf c)}~With imperfections $\eps_J=10^{-2} , \, \eps_B=0$ 
	averaged over $N_{av}=10^2$ realizations.}
  	\label{fig:ft}
  	\end{center}
 \end{figure}
The state (\ref{icond}) will evolve accordingly to the dynamics dictated by 
Eq.~(\ref{spinchain}). Since the Hamiltonian commutes with the total spin component 
along the $z$ direction, the relevant sector of the Hilbert space is 
spanned by the states
\begin{eqnarray}
	| {\boldsymbol j}\rangle \equiv |0,0, 
	 \cdots 0, 1,0, \cdots 0 \rangle \;, 
\label{vect}
\end{eqnarray}
which for $j=1,\cdots,N$ represents a state of the chain where the $j$th spin 
is prepared in $|1 \rangle$ and the other $N-1$ ones
in $|0\rangle$. 
The global state of the chain at time $t$ is
\begin{eqnarray}
	|\Psi(t)\rangle = \cos\, \vartheta | {\boldsymbol 0}\rangle + \sin\, \vartheta
	e^{\imath \varphi}\sum_{j=1}^N f_{j}(t) |{\boldsymbol j}\rangle \;,
\label{OUTPUT}
\end{eqnarray}
where $|{\boldsymbol 0}\rangle$ is the chain state with all the spins
in $|0 \rangle$ and where (we set $\hbar=1$)
\begin{eqnarray}
	f_{j}(t)\equiv
	\langle {\boldsymbol j} | e^{-i H t}|
	{\boldsymbol 1}\rangle \label{effe}\;.
\label{amplitude}
\end{eqnarray}
The accuracy of the state transfer is determined through the analysis of the fidelity
$$
	\fid (t, \vartheta, \varphi)=\bra {\Psi_0(\vartheta,\varphi)}
	\rho_N (t) \ket {\Psi_0(\vartheta,\varphi)} 
$$ 
where $\rho_N (t)$ is the reduced density matrix
of the $N$th spin at time $t$. We consider the fidelity averaged over the
initial state $\ket {\Psi_0}$ distributed uniformly over the Bloch
sphere \cite{bose},
\beq
\fid(t)= \langle \fid (t, \vartheta,
\varphi)\rangle_{\vartheta,\varphi} =  \frac{|f_{N}|}{3}+\frac{|f_{N}|^2}{6}+\frac{1}{2}.
\label{fidf}
\eeq
We ignore the phase of $f_N$ as it can be gauged away by a proper
choice of the external field. 
Eventually, we will average over different disorder realizations, that is, 
\beq
\overline{\fid}(t)= \langle \fid(t) \rangle_{\mathcal{D}}
\label{avfid}
\eeq
where $\langle .\rangle_{\mathcal{D}} $ stands for the average over different imperfection
configurations.
In Ref.~\cite{christandl03} it has been shown that after a time  $t_n=~(2n+1)\pi/4J$ 
($n$ integer) the state of the left-most spin is transferred exactly to the 
right-most spin. This is due to the fact that the 
Hamiltonian (\ref{icond}) can be viewed as that of a pseudospin $\vec S=(N-1)/2$ 
that precesses in a constant magnetic field. The  state transmission 
is equivalent to a $\pi$ rotation of the spin.
\begin{figure}[htbp]
  	\begin{center}
    	\includegraphics[scale=0.35]{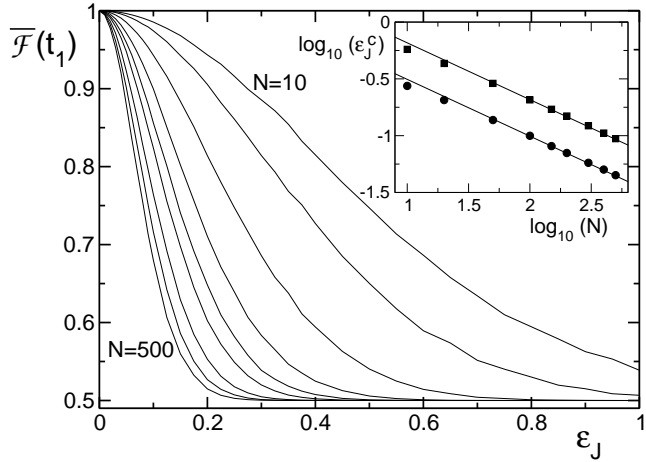}     
 	\caption{Averaged fidelity at time $t_1$ as a function of the disorder
   	$\eps_J$ for different spin-chain lengths and  $\eps_B=0$, $N_{av}=10^3$.
	From right to left $N=10, \, 20, \, 50, \, 100, \, 150, \, 200, \, 300,
	\, 400, \, 500$. Inset: $\eps_J^c$ as a function of $N$ obtained from
   	the condition $\overline{\fid}(t_1)=0.9$ (circles) and  $\overline{\fid}(t_1)=0.7$
   	(squares). Straight lines are proportional to $N^{-0.5}$.
	Here and in the following figures the logarithms are decimal.}
  	\label{fig:fepsj}
  	\end{center}
 \end{figure}
In order to analyze the robustness of this protocol to static imperfections,
we model their effects by adding to the Hamiltonian  a 
random perturbation both in the exchange couplings and in the local variations 
of the magnetic field. The coefficients in Eq.(\ref{spinchain}) are replaced 
with the new values
$$
	 B_k \rightarrow b_k,\,\,\,\,\,J_k \rightarrow J_k (1+\delta_{k})
$$ 
where $\delta_k$ and $b_k$  are random variables with uniform distribution 
in the intervals $\delta_{k} \in [- \eps_J , \eps_J ]$ and
$b_k \in [- \eps_B , \eps_B ]$. The results presented in this paper are 
obtained by averaging over $N_{av}$ different disorder
realizations. 

\section{Stability of the communication in a disordered chain}
\label{sec:stab}

We numerically solve the Schr\"odinger equation for the dynamical evolution and 
compute the fidelity of the rightmost spin with respect to the input state. 
In Fig.~\ref{fig:ft} we plot typical results of this evolution both for the
ideal case (Fig.~\ref{fig:ft}A) and in presence of imperfections
(Fig.~\ref{fig:ft}B.). Figure \ref{fig:ft}C is the result of an
average over different disorder realizations.
In the presence of disorder the simple periodicity of the fidelity oscillation is
lost. Moreover, the maximal value of the fidelity is less than unity (it is
reached at slightly different time intervals as compared to the ideal
case). Thus the optimal time for state transfer should be inferred
for each experimental sample.  
The original (in the ideal case) periodicity of the signal is
recovered averaging over different disorder realizations, however, the 
maxima are progressively suppressed on increasing time.
Therefore the optimal state transfer, in presence of 
imperfections, is obtained in correspondence of the first peak at time $t_1=\pi/4J$. 

In this section we concentrate on the dependence of the optimal
fidelity ($ \overline{\fid}$ at 
time $t_1$) as a function of static imperfection strength and of the chain length.

\begin{figure}[htbp]
  	\begin{center}
     	\includegraphics[scale=0.35]{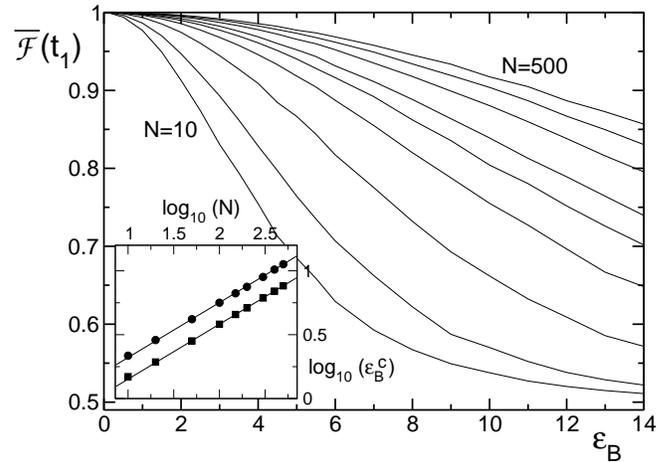}
 	\caption{Averaged fidelity at time $t_1$ as a function
	of magnetic field disorder $\eps_B$ for different spin-chain lengths
	and  $\eps_J=0$, $N_{av}=10^3$.
	From left to right $N=10, \, 20, \, 50, \, 100, \, 150, \, 200, \, 300,
	\, 400, \, 500$.  
	Inset: $\eps_B^c$ as a function of $N$ obtained from	
	the condition $\overline{\fid}(t_1)=0.9$ (circles) and
	$\overline{\fid}(t_1)=0.95$ (squares).
 	Straight lines  are proportional to $N^{0.43}$.}
  	\label{fig:fepsb}
  	\end{center}
 \end{figure}
\begin{figure}[htbp]
  \begin{center}
     \includegraphics[scale=0.8]{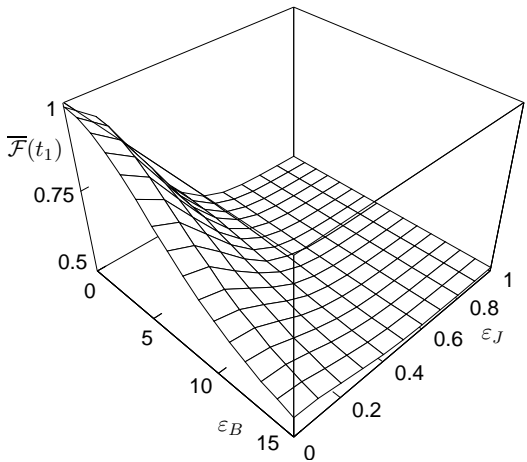}
  \caption{Average fidelity at time $t_1$ as a function of the amplitudes
of disorder  $\eps_J$, $\eps_B$ for a $N=50$ spin network, $N_{av}=100$.}
  \label{fig:febj}
  \end{center}
 \end{figure}

In Fig.~\ref{fig:fepsj} we report the fidelity as a function of $\eps_J$ 
for different chain lengths assuming, for the moment, that there is no disorder
in the local field ($\eps_B=0$). The opposite situation, with disordered local magnetic
field ($\eps_B \ne 0$) and ideal nearest neighbor interaction ($\eps_J=0$) is
shown in Fig.~\ref{fig:fepsb}. These sources of disorder lead to a
striking different behavior. While in the first case the error
introduced by the imperfections increases
with $N$, the effect of the disorder on local magnetic field decreases, 
becoming less effective on increasing the chain length. For completeness 
we show the case where both $\eps_B$ and $\eps_J$ are different from
zero in Fig.~\ref{fig:febj}.
The fact that the two effects are almost independent can be traced 
back to the fact that the we are working in the sector with one spin up. 

The behavior of the fidelity obeys a simple scaling law. We verified numerically that 
the fidelity scales as
\beq
	\overline{\fid} (t_1) = 
	\frac{1}{2}(1+e^{-\kappa_J N \eps_J^2-\kappa_B  \eps_B^2 /N})
\label{fidfit}
\eeq
where $\kappa_J \sim 0.2$ and $\kappa_B \sim 0.7$. The constants
$\kappa_J$ ($\kappa_B$) have been obtained from the dependence, as a function of
$N$, of the value $\eps_J^c$  ($\eps_B^c$) at which the fidelity reaches a given threshold
value (see the insets of Figs.~\ref{fig:fepsj} and \ref{fig:fepsb}).

The scaling given in Eq.~(\ref{fidfit}) can be justified in the limit of very small 
disorder by means of perturbation theory. In the limit  $\eps_J t, \eps_B t \ll 1 $,  
the fidelity reads
\barr
\overline{\fid}(t) &\approx& 1 - \frac{\varepsilon_B^2}{3}
\sum_{k=1}^N \Big( 2\, \Re e [D_{k,k}(t)] - C_k^2(t) \Big)/3 \nonumber \\
&-& \frac{\varepsilon_J^2}{3} 
\sum_{k=1}^N \Big( 2\, \Re e [F_{k,k}(t)]- E_k^2(t) \Big)/3
\label{pert}
\earr
The coefficients $C_k, \, D_{k,k}, E_k$, and $F_{k,k}$ as well as the details of the
calculation are given in the appendix \ref{app1}.
The disorder in the local magnetic field averages out in the limit of infinite spin chains.
In view of the little effect of random fields on the quantum communication over 
long chains, from now on we will consider only the effect of disordered exchange 
coupling between spins.

\begin{figure}[htbp]
  \begin{center}
    \includegraphics[scale=0.35]{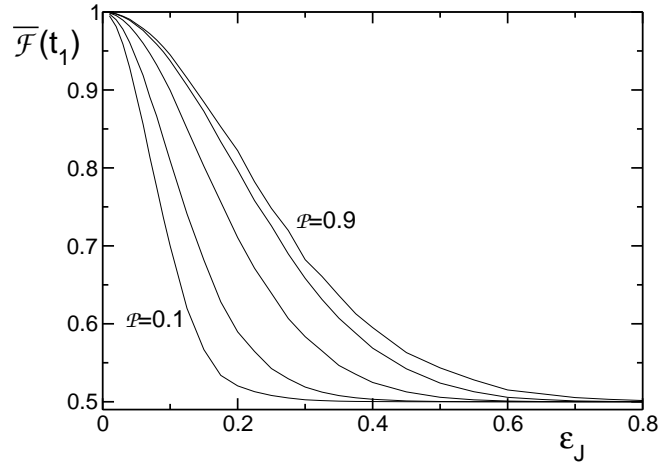}
    \caption{Fidelity at time $t_1$ as a function
      of $\eps_{J}$ with $N=100$, $N_{av}=200$ and from left to right
      $\mathcal{P}=0.1, \, 0.25, \, 0.5, \, 0.75, \, 0.9$.}
    \label{fig:corr}
  \end{center}
\end{figure}

The presence of spatial correlation in the disorder is a concrete possibility 
in experimental realizations  of this protocol, as, for example, with Josephson-junction 
chains~\cite{romito,paternostro}.
We model correlated disorder as follows: The sign of any single
$\delta_k$, the error on the $k$th coupling, is correlated with 
the previous one following the rule:
\barr
& &	\delta_{i}\delta_{i-1} > 0 ~~ \textrm{\small {with
    probability}} ~ \mathcal{P},  \nonumber \\
& &      \delta_{i}\delta_{i-1} < 0 ~~
	 \textrm{\small otherwise}. 
\label{correl}
\earr
The correlations introduced in Eq.(\ref{correl}) result in a  
perfect correlation (anticorrelation) in the signs between the fluctuations among nearest 
neighbors if $\mathcal{P}=1$ ($\mathcal{P}=0$). Uncorrelated disorder
is recovered for $\mathcal{P}=0.5$. In Fig.~\ref{fig:corr} the fidelity 
as a function of $\eps_J$ ($\eps_B=0$) for different values of 
$\mathcal{P}$ is plotted. Notice that the fidelity decay is a monotonic function
of $\mathcal{P}$. Anticorrelated disorder is more dangerous than
correlated one. 

Within the model of disorder studied in this work, strong 
fluctuations of the exchange couplings lead to a degradation of the
signal while the same protocol is not very sensitive (especially for 
long chains) to fluctuations in the local magnetic fields.

\section{Level spacing statistics}\label{sec:lss}

The behavior of the fidelity is essentially dictated by the 
time dependence of the amplitude $f_{N}$ defined in Eq.~(\ref{amplitude}).
A deeper insight of its characteristics in disordered chains can 
be understood by analyzing  the statistics of the level spacing  of
the spin-chain Hamiltonian in presence of disorder. The level spacing 
statistics $P(s)$ is widely used to study complex many-body 
systems~\cite{lss1} and quantum 
systems with classically chaotic counterparts~\cite{lss2} in the
framework of random matrix theory~\cite{rmt}.
The distribution $P(s)ds$ gives the probability that the energy difference
between two adjacent levels (normalized to the average level spacing)
belongs to the interval $[s,s+ds]$. The Hamiltonian (\ref{spinchain}) 
is  a tridiagonal matrix and thus it is
not a random matrix, however, we will show that this analysis helps in understanding 
the behavior of the disordered chain.  The level spacing
statistics can still be used to characterize the crossover that static
imperfections induce in the spectrum of the Hamiltonian.
\begin{figure}[htbp]
  \begin{center}
     \includegraphics[scale=0.35]{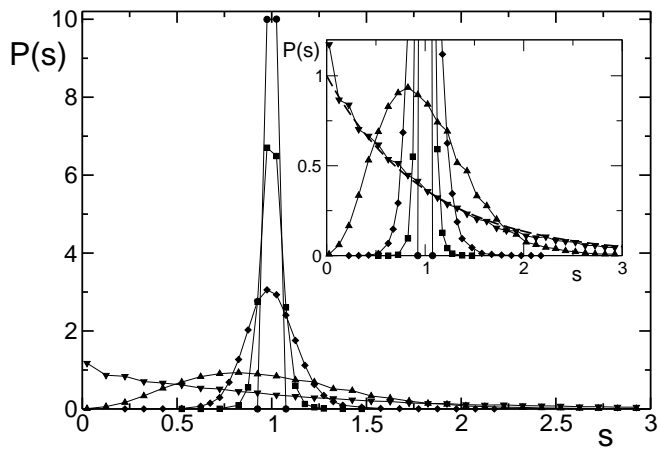}
     \caption{Level spacing statistics $P(s)$ for $N=100,
       \epsilon_B=0$, $N_{av}=10^3$  and different values of $\epsilon_J$:
       $\epsilon_J=10^{-3}$ (circles), $\epsilon_J=2 \times 10^{-2}$ (squares),
       $\, \epsilon_J=5 \times 10^{-2}$ (diamonds), $\,\epsilon_J=2
       \times 10^{-1}$ (triangles up), $\, \epsilon_J=1$ (triangles
       down). 
       Inset: magnification of the same figure around $s=1$. The dashed line corresponds
       to the Poissonian $P_P (s)$.}
     \label{fig:levels}
  \end{center}
 \end{figure}

\begin{figure}[htbp]
  \begin{center}
     \includegraphics[scale=0.35]{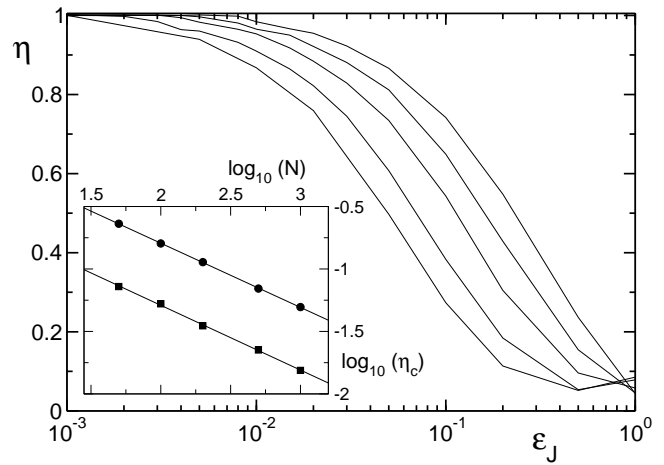}
  \caption{The parameter $\eta$ as a function of the strength of the
    static imperfections $\epsilon_J$. Different curves correspond to different
spin chain length: from right to left $N=50$, $\, N=100$, $\, N=200$,
    $\, N=500$. We averaged over $N_{av}$ disorder realizations with
    $N_{av}=10^4$.
Inset: scaling of the parameter
$\eta_c$ as a function of the chain length $N$ obtained from
the condition $\eta=0.5$ (circles) and  $\eta=0.8$ (squares). 
Straight lines are proportional to $N^{-0.5}$.}
  \label{fig:eta}
  \end{center}
 \end{figure}

The Hamiltonian 
$H$ is a tridiagonal matrix with zero entries on the diagonal, and
$H_{k,k+1} = H_{k+1,k} = \lambda \sqrt{k (N- k)}$
where $N$ is the chain length and $\lambda$ a constant.
Without any perturbation ($\epsilon_J = \epsilon_B = 0$) the energy 
levels are then equally spaced, while in presence of strong random perturbations 
($\vert \epsilon_J \vert \sim 1$) its eigenvalues are completely uncorrelated.
This crossover is detected by the level spacing statistics. It 
changes from a delta function to a Poisson
distribution given by the formula
\begin{eqnarray}
	P_D (s)&=&\delta(s-1) \,\,\,\,\,\, \mbox{no disorder}\\ 
	P_P (s)& =& \exp (-s) \,\,\,\,\,\, \mbox{strong disorder} .
\end{eqnarray}
Figure~\ref{fig:levels} shows this crossover: $P(s)$ changes from
one limiting case to the other as a function of static imperfection
strength. This crossover can be quantitatively characterized by 
the parameter: 
\begin{equation}
	\eta = \frac{\int_0^1 \vert P(s) - P_P (s) \vert ds}
	{\int_0^1 \vert P_D (s) - P_P (s) \vert ds},
\end{equation}
which varies from $\eta = 1$ in the case of a delta function to
$\eta=0$ for a Poisson distribution~\cite{core}.
In Fig.~\ref{fig:eta} we show the dependence of $\eta$ on the strength
of the perturbation. The crossover starts at $\epsilon_J \sim 10^{-3}
- 10^{-2}$ depending on the length of the chain.
In the inset of Fig.~\ref{fig:eta} we report the dependence of the
imperfection strength $\eta_c$ at which the parameter $\eta$ reaches a
given constant value ($\eta=0.5, 0.8$). The threshold $\eta_c$ drops with the spin
length as
$$
\eta_c \sim N^{-0.5}.
$$
Thus it follows the same law found in the previous section regarding the
fidelity of the state transfer.

In the next section, we show that the same crossover is reflected by the fidelity
time series with the appearance of a fractal behavior. 

\section{Fractal dimension of the fidelity}\label{sec:frac}

An interesting consequence of the modification of the spectrum, and 
hence of the fidelity, in presence of static imperfections 
emerges in the time dependence of the fidelity. In this section we 
will not look for the optimal time for the state transfer but rather 
analyze its behavior as a function of time. It appears that the 
time signal of the fidelity has a fractal behavior.
In order to measure the fractal dimension of the signal we used the modified
box counting algorithm~\cite{ketzmerick}.
In the standard box counting algorithm the fractal
dimension $D$ of the signal is obtained by covering the data with a
grid of square boxes of size $L^2$. The number $M(L)$ of boxes
needed to cover the curve is recorded as a function of the box
size $L$. The (fractal) dimension $D$ of the curve is then defined as 
\beq
	D = - \lim_{L \to 0} \log_L M(L). 
\label{fractal}
\eeq
One finds $D=1$ for a straight line, while $D=2$
for a periodic curve. Indeed, for times much larger than the period, a
periodic curve covers uniformly a rectangular region.  Any given value of $D$ in
between of these integer values is a signal of the fractality of the curve. 
The modified algorithm of Ref.\cite{ketzmerick} follows the same lines but 
uses rectangular boxes of size $ L \times \Delta_i$ ($\Delta_i$ is the largest 
excursion  of the curve in the region $L$). Then, the number 
\beq
	M(L) = \frac{\sum_i  \Delta_i}{L} 
\label{emme}
\eeq 
is computed (the time boxes $L$ are expressed in units of the exchange
coupling $J$). For any curve a region of box lengths $L_{min} 
< L < L_{max}$ exists where $M \propto L^D$. Outside this region one
either finds $D=1$ or $D=2$: The first equality ($D=1$) holds for
$L<L_{min}$ and it is due to the coarse grain artificially introduced
by any numerical simulations. The second one ($D=2$) is obtained for
$L>L_{max}$ and it is due to the finite length of the analyzed time
series. The boundaries $L_{min},L_{max}$ have to be chosen properly for
any time series. 
\begin{figure}[htbp]
  \begin{center}
     \includegraphics[scale=0.33]{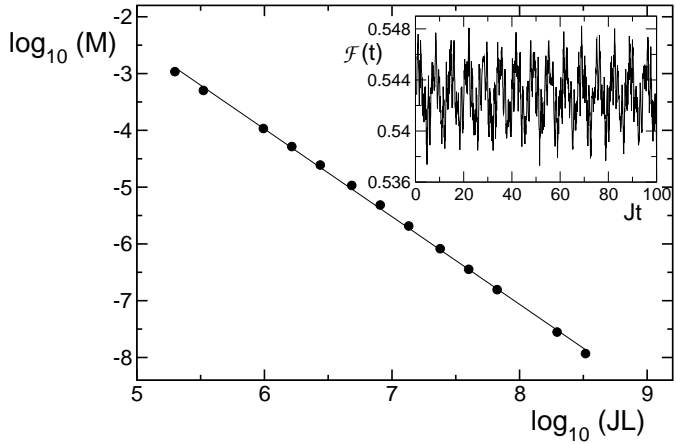}
  \caption{$M$ as a function of the interval $J L$. 
A numerical fit gives a fractal dimension $D=1.52$. 
Inset: Temporal evolution of the fidelity up to time $T=10^4/J$ in 
the presence of disorder for $\eps_J=0.26$, $\eps_B=0$, $N_{av}=~1$, $N=500$. }
  \label{fig:exmplfrac}
  \end{center}
 \end{figure}

\begin{figure}[thbp]
  \begin{center}
     \includegraphics[scale=0.35]{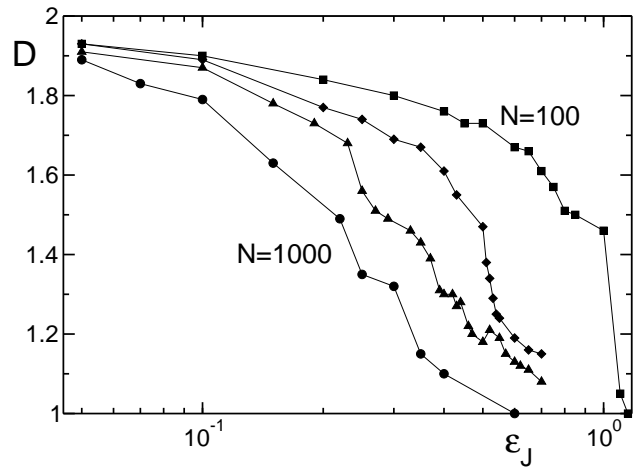}
     \caption{Fractal dimension $D$ of the signal $ \fid (t)$ as a function
       of the perturbation strength $\eps_J$ for $\eps_B=0, \, N_{av}=1$
       and, from right to left, $N=100, \, N=200, \, N=500, \, N=1000$. 
       The error on the fractal dimension is of the order of three
       percent. For $\eps_J < 5 \times 10^{-2}$ the error on the fractal
       dimensions $D$ increases significantly as $L_{min} \lesssim L_{max}$.}
     \label{fig:epsfrac}
  \end{center}
\end{figure}

\begin{figure}[thbp]
  \begin{center}
     \includegraphics[scale=0.35]{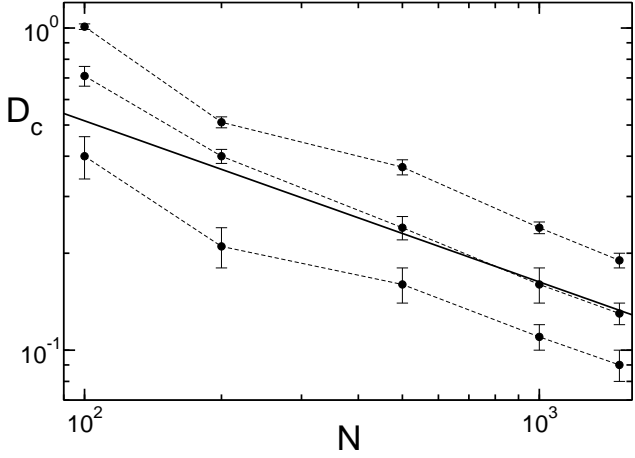}
     \caption{Scaling parameter $D_c$ as a function of chain
       length $N$ from the condition $D=1.76, \, 1.6, \, 1.4$ (from bottom to
       the top) from
       Fig.~\ref{fig:epsfrac}. The full straight line is proportional to $N^{-0.5}$.}
     \label{fig:diminset}
  \end{center}
\end{figure}

We apply the modified algorithm to the signal of the fidelity
for a single realization of disorder after a transient regime needed
to reach the average value of $\fid=0.5$: The inset of
Fig.~\ref{fig:exmplfrac} shows the typical fluctuating signal we
analyzed while Fig.~\ref{fig:exmplfrac} shows the numerically computed
function $M(L)$ which gives a fractal dimension $D=1.52$. 
It is natural to investigate the dependence of the
fractal dimension with the static imperfection strengths: the results of
numerical simulations are given in Fig.~\ref{fig:epsfrac}. The
curve changes gradually its dimension from $D \approx 2$ (periodic
curve) to $D=1$ for very large imperfection strengths. This last
result is due to the fact that for very large disorder the fidelity
drops almost immediately to $0.5$ corresponding to a complete loss of
the initial state information: The fidelity remains then constant,
characterized by dimension $D=1$.  However, the most general
situation in presence of static imperfections is a fidelity with fractal dimension: 
defining, as before, a threshold of disorder strength $D_c$ at 
which the fidelity has a given fractal dimension (between two and
one), we find that this threshold drops as
$$
D_c \sim N^{-0.5}.
$$ 
This behavior is shown in Fig.~\ref{fig:diminset} and follows exactly
the same scaling as the parameters $\eta_c$ and $\epsilon_J^c$.

\section{conclusions}
We have shown that static imperfections in a modulated spin chain
destroy, above a given threshold, the transmission of quantum states if
performed following the protocols presented in Ref.~\cite{christandl03}. 
We characterize the effects of static imperfections by means of the
fidelity of the state transmission. This transition is reflected 
in the changing of the level spacing statistics (from delta-correlated
to completed uncorrelated) and in the behavior of the fidelity time
series: the perfect state transfer is characterized by a periodic
fidelity with integer fractal dimension while beyond the
critical threshold it is described by a fractal dimension. 
We characterize these crossovers by 
analyzing $\epsilon_J^c, \eta_c, D_c$: the imperfection strength needed to reach this 
value defines a critical threshold. 
The three distinct critical thresholds follow the same scaling as 
a function of the chain length and imperfection strength, 
independently from the critical value chosen.
This common behavior reflects the profound changes in the quantum 
system induced by the presence of static imperfections.  
The threshold drops as the square root of the chain length: 
this is a behavior similar to the one found in Ref.~\cite{benenti} in a different system 
where it was a consequence of the two body nature of the interactions. 
Here, the dependence is mainly due to the fact that the system 
is confined to the subspace of one excitation. The conclusion of this
analysis is that it is possible, at least in 
principle, to tolerate or correct the errors introduced by static 
imperfection. 

\acknowledgments
We acknowledge very fruitful discussions with V. Giovannetti and A. 
Romito. This work was supported by the European Community under
contracts  IST-SQUBIT2, RTN-NANO, and MIUR-Firb.

\appendix
\section{Perturbation Theory}
\label{app1}
We are interested in evaluating the fidelity (\ref{avfid}) averaged
over different disorder realizations. Equation (\ref{fidf}) shows that
it depends on the matrix element 
\barr
f_N(t) &=&\langle \boldsymbol{N} | e^{-\imath (H+H_I) t} | \boldsymbol{1}
\rangle \label{pertex} \\
&=& 
\langle \boldsymbol{N} | e^{-\imath H t} 
\mathcal{T}\left[\exp{\left(-\imath
\int_0^{t} dt \, e^{\imath H t} H_I e^{-\imath H t}\right)}\right] 
| \boldsymbol{1} \rangle \nonumber \\
&=& 1 + \mathcal{O}(H_I) + \mathcal{O}(H_I^2) \nonumber
\earr
where $\mathcal{T}$ is the time ordered product, $\hbar=1$, 
and $H_I$ is the part of the Hamiltonian that 
describes the static imperfections $b_k, \delta_k$.
We first consider the case where $\delta_k=0$, that is, only random
local magnetic fields are present. 
We develop the time ordered product up to the second order in
$H_I$. The first order term reads
\barr
\mathcal{O}(H_I) &=& - \imath \int_0^t dt \langle \boldsymbol{1} | e^{\imath H t} H_I
e^{-\imath H t} | \boldsymbol{1} \rangle  \\
&=&- \imath \sum_{\ell=1}^N b_\ell \int_0^t dt (1-2 |U_\ell^1(t)|^2)
\equiv - \imath \sum_{\ell=1}^N b_\ell C_\ell(t), \nonumber
\earr
where $U_\ell^k(t) \equiv  \langle \boldsymbol{\ell} |e^{-\imath H t}
|\boldsymbol{k} \rangle$ \, \cite{feynman}. The second order is given by
\barr
\mathcal{O}(H_I^2) &=& - \int_0^t \int_0^t  dt dt'  
\langle \boldsymbol{1} | e^{\imath H t} H_I e^{-\imath H (t-t')} H_I
e^{-\imath H t'}|\boldsymbol{1} \rangle  \nonumber 
\earr
\barr
&=& - \sum_{\ell=1}^N \sum_{m=1}^N b_\ell b_m \int_0^t \int_0^t  dt \,
dt' 
\left[ 1- 2 \abs{U_m^1(t)}^2 \right. \nonumber \\
&- & \left.  2 \abs{U_\ell^1(t')}^2 + 4 U_m^{1*}(t) 
  U_\ell^1(t') \sum_{k=1}^N U_m^{k}(t) U_k^{\ell *}(t')  \right]
\nonumber \\
&\equiv& - \sum_{\ell=1}^N \sum_{m=1}^N b_\ell b_m D_{\ell, m}(t).
\earr
The fidelity (\ref{avfid}) is given by the average over
different disorder realization of the coefficient $f_N(t)$ and of its
modulus square:
\barr
\overline{\fid(t)} &=& \frac{1}{2}+\langle \frac{|f_{N}(t)|}{3} +\frac{|f_{N}(t)|^2}{6}
\rangle_{\mathcal{D}} \\
&\approx& 1 - \frac{\varepsilon_B^2}{3} \sum_{k=1}^N \Big( 2\, \Re e [D_{k,k}(t)]
- C_k^2(t) \Big) /3 \nonumber.
\earr
The case for $\delta_k \ne 0$ is obtained following the same steps 
and with a final result of
\barr
\overline{\fid(t)} &=& 1 - \frac{\varepsilon_B^2}{3} \sum_{k=1}^N \Big( 2\, \Re e [D_{k,k}(t)]
- C_k^2(t) \Big) /3 + \nonumber \label{pertf}\\
& - & \frac{\varepsilon_J^2}{3} \sum_{k=1}^N \Big( 2\, \Re e [F_{k,k}(t)]
- E_k^2(t)\Big) /3 ,
\earr
where the coefficients $E_k, F_{m,\ell}$ are given by
\barr
E_\ell &=& 4 \int^t_0 dt \, \Re e \, [U_\ell^1(t) U_{\ell+1}^{1 *}(t)] \, , \\
F_{m,\ell} &=& 4 \int^t_0 \int^t_0 dt \, dt' \sum_{k=1}^N 
\left(U_m^{1 *}(t) U_{k}^{m+1}(t) \right. + \nonumber \\
&&\left. U_{m+1}^{1 *}(t) U_{k}^{m}(t) \right) 
\left(U_\ell^{1 *}(t') U_{k}^{\ell+1 *}(t') \right. + \nonumber \\ 
&&\left. U_{\ell+1}^{1 *}(t') U_{k}^{\ell *}(t') \right). 
\nonumber 
\earr
The effects of the two different kind of perturbations in
Eq.~(\ref{pertf}), the local magnetic $b_\ell$ fields and 
the couplings $\delta_k$, are decoupled because they fluctuate
independently from each other. 



\begin{thebibliography}{99}
\bibitem{gisin}
	N. Gisin, G. Ribordy, W. Tittel, and H. Zbinden,
        Rev. Mod. Phys. {\bf 74}, 145  (2002).
\bibitem{teleportation}
	C. H. Bennett, G. Brassard, C. Cr\'epeau, R. Jozsa, 
	A. Peres, and W. K. Wootters, Phys. Rev. Lett.
	{\bf 70}, 1895 (1993).	
\bibitem{tian}
	L. Tian, P. Rabl, R. Blatt, and P. Zoller, 
	Phys. Rev. Lett. {\bf 92}, 247902 (2004).
\bibitem{bose} 
	S. Bose, Phys. Rev. Lett. {\bf 91}, 207901 (2003).
\bibitem{giovannetti}
	V. Giovannetti, and R. Fazio,
	Phys. Rev. A {\bf 71}, 032314 (2005).
\bibitem{Li}
        Y. Li, T. Shi, B. Chen, Z. Song, and C. P. Sun,
	Phys. Rev. A {\bf 71}, 022301 (2005).
\bibitem{romito} 
	A. Romito, R. Fazio, and C. Bruder,
	Phys. Rev. B {\bf 71}, 100501(R) (2005).
\bibitem{paternostro} 
	M. Paternostro, G. M. Palma, M.S. Kim, and G. Falci,
  	Phys. Rev. A {\bf 71}, 042311 (2005).
\bibitem{christandl03} 
	M. Christandl, N. Datta, A. Ekert, and A. J. Landahl,
	Phys. Rev. Lett. {\bf 92}, 187902 (2004).
\bibitem{veer}  
	F. Verstraete, M. A. Mart\'in-Delgado, and
  	J. I. Cirac, Phys. Rev. Lett. {\bf 92}, 087201 (2004).
\bibitem{osborne03}
        T. J. Osborne, and N. Linden, Phys. Rev. A {\bf 69}, 052315 (2004).
\bibitem{burgarth} 
	D. Burgarth, and S. Bose, Phys. Rev. A {\bf 71}, 052315 (2005).
	D. Burgarth, V. Giovannetti, and S. Bose, quant-ph/0410175.
\bibitem{core} 
	B. Georgeot, and D. L. Shepelyansky, Phys. Rev. E {\bf 62}, 6366 (2000).
\bibitem{benenti} 
	G. Benenti, G. Casati, S. Montangero, and D. L. Shepelyansky,
  	Phys. Rev. Lett. {\bf 87}, 227901 (2001).
\bibitem{Zhang}
        J. Zhang, G. L. Long, W. Zhang, Z. Deng, W. Liu, and Z. Lu,
        quant-ph/0503199.
\bibitem{rmt} 
        T. Guhr, A. M\"uller-Groeling, and H. A. Weidenm\"uller, 
        Phys. Rep. {\bf 299}, 190 (1998).
\bibitem{berry} 
	M. V. Berry, J. Phys. A {\bf 29}, 6617 (1996).
\bibitem{amanatidis} 
	E. J. Amanatidis, D. E. Katsanos, and S. N. Evangelou,
  	Phys. Rev. B {\bf 69}, 195107 (2004).
\bibitem{lss1} 
	F. Haake, {\it Quantum Signature of chaos} (Springer-Verlag,
	New York, 1991). 
\bibitem{lss2} 
  	F. Izrailev, Phys. Rep. {\bf 196}, 299 (1990). 
\bibitem{ketzmerick} 
        A. S. Sachrajda, R. Ketzmerick, C. Gould, Y. Feng, P. J. Kelly, A. Delage,
        and Z. Wasilewski, Phys. Rev. Lett. {\bf 80}, 1948 (1998).
\bibitem{feynman}
        The matrix elements $U_\ell^k(t)$ are well known, and can be found in:
        R. P. Feynman, R. B. Leighton, and M. Sands,
        {\it Feynman lectures on Physics} (Addison-Wesley, Reading, MA, 1965), Vol.~3.
\end{thebibliography}
\end{document}